\documentclass[sigconf]{acmart}
\usepackage{comment}
\usepackage{booktabs, tabularx, array}
\usepackage[most]{tcolorbox}
\usepackage[table]{xcolor}
\usepackage{xcolor}
\usepackage{multirow}
\usetikzlibrary{shadows}
\usepackage{graphicx}
\usepackage{subcaption}
\usepackage{array,tabularx,multirow,colortbl,xcolor}
\usepackage{pifont}
\usepackage{booktabs,tabularx,colortbl,fontawesome5}
\definecolor{SOneHeader}{RGB}{235,235,235}
\definecolor{SOneGroup}{RGB}{246,246,246}
\definecolor{SOneAccent}{RGB}{135,31,37}

\definecolor{SOneAccent}{RGB}{135,31,37}
\definecolor{SOneFindingBG}{RGB}{248,240,240}

\newtcolorbox{findingbox}[1]{
  enhanced,
  breakable=false,
  colback=SOneFindingBG,
  colframe=SOneAccent,
  boxrule=0.5pt,
  sharp corners,
  left=5pt,
  right=5pt,
  top=4pt,
  bottom=4pt,
  before upper={%
    \textcolor{SOneAccent}{\bfseries Finding #1.}\enspace
  },
  drop shadow={
    opacity=0.12,
    shadow xshift=0.5mm,
    shadow yshift=-0.5mm
  },
  before skip=6pt,
  after skip=6pt
}
\newcommand{\finding}[2]{%
  \begin{findingbox}{#1}%
  #2%
  \end{findingbox}%
}
\usepackage{enumitem}
\definecolor{SOneUse}{RGB}{36,87,149}
\definecolor{SOneInput}{RGB}{20,118,109}
\definecolor{SOneKeys}{RGB}{119,67,143}
\definecolor{SOneCode}{RGB}{168,100,21}

\definecolor{kwhead}{HTML}{2C3E50}
\definecolor{kwsearch}{HTML}{2E86C1}
\definecolor{kwcode}{HTML}{CA6F1E}
\definecolor{kwstripe}{HTML}{F4F6F7}
\newtcbox{\kws}{on line, nobeforeafter, boxsep=0pt, left=2pt, right=2pt, top=1.2pt, bottom=1.2pt,
  arc=1.6pt, boxrule=0.35pt, colback=kwsearch!7, colframe=kwsearch!45, fontupper=\ttfamily\scriptsize}
\newtcbox{\kwc}{on line, nobeforeafter, boxsep=0pt, left=2pt, right=2pt, top=1.2pt, bottom=1.2pt,
  arc=1.6pt, boxrule=0.35pt, colback=kwcode!8, colframe=kwcode!50, fontupper=\ttfamily\scriptsize}

\definecolor{rqbg}{RGB}{248,248,248}
\definecolor{rqframe}{RGB}{95,95,95}

\newtcolorbox{rqbox}{
    enhanced,
    colback=rqbg,
    colframe=rqframe,
    boxrule=0.8pt,
    arc=2.2mm,
    outer arc=2.2mm,
    left=2.2mm,
    right=2.2mm,
    top=1.7mm,
    bottom=1.7mm,
    boxsep=0pt,
    before skip=6pt,
    after skip=6pt,
    drop shadow={
        opacity=0.18,
        shadow xshift=0.7mm,
        shadow yshift=-0.7mm
    }
}

\newtcolorbox{developerquote}[1]{
    enhanced,
    breakable,
    colback=black!2,
    colbacktitle=black!8,
    colframe=black!65,
    coltitle=black,
    boxrule=0.5pt,
    arc=1pt,
    fonttitle=\footnotesize\bfseries,
    title={#1},
    titlerule=0.4pt,
    left=7pt,
    right=7pt,
    top=6pt,
    bottom=6pt,
    before skip=5pt,
    after skip=5pt,
    drop shadow={
        opacity=0.18,
        shadow xshift=0.7mm,
        shadow yshift=-0.7mm
    }
}

\AtBeginDocument{%
  }

\setcopyright{acmlicensed}
\copyrightyear{2018}
\acmYear{2018}
\acmDOI{XXXXXXX.XXXXXXX}
\acmConference[Conference acronym 'XX]{Make sure to enter the correct
  conference title from your rights confirmation email}{June 03-05,
  2018}{Woodstock, NY}
\acmISBN{978-1-4503-XXXX-X/2018/06}
\makeatletter
\renewcommand{\@ACM@checkaffil}{}

\AtBeginDocument{%
    \fancyhead[L]{\ACM@linecountL}
    \fancyhead[R]{\ACM@linecountR}
}

\let\@origfootnotetextcopyrightpermission\footnotetextcopyrightpermission
\renewcommand\footnotetextcopyrightpermission[1]{
  }
\makeatother
\newcommand{\bo}[1]{#1} 
\begin{document}

\title{Fast Decisions, Hidden Risks: Measuring the Security Portrait of System One Model Applications from Jev and Beyond}

\author{Bo Li$^{1,2}$, Yue Yang$^{2}$, Xin Zheng$^{3}$,
David Lemphers$^{2}$, Shirui Pan$^{1}$,
Xingliang Yuan$^{4}$, Yujin Huang$^{5,\dagger}$}

\affiliation{
\institution{
$^{1}$Griffith University, Australia
$^{2}$Maincode, Australia
$^{3}$RMIT University, Australia\\
$^{4}$The University of Melbourne, Australia
$^{5}$Adelaide University, Australia
}}

\affiliation{
\institution{
$^{1}$\{bo.li, s.pan\}@griffith.edu.au
$^{2}$\{yue, dave\}@maincode.com
$^{3}$xin.zheng2@rmit.edu.au\\
$^{4}$xingliang.yuan@unimelb.edu.au
$^{5}$yujin.huang@adelaide.edu.au
}}



\providecommand{\finding}[2]{%
  \par\smallskip\noindent\fbox{\parbox{\dimexpr\linewidth-2\fboxsep-2\fboxrule\relax}{\textbf{Finding #1:} #2}}\par\smallskip}

\begin{abstract}

System One models are an emerging class of AI models that provide fast, structured judgments for software applications in the form of typed decisions with associated probabilities rather than free-form text.
Since the release of Jev in September 2026, developers have rapidly adopted System One models to gate actions, route requests, and screen content.
However, security failures have already been reported, and the vulnerabilities arising from the integration of these models into real-world applications remain poorly understood.
In this paper, we present the first systematic measurement study of security vulnerabilities in System One model applications, characterizing the vulnerability landscape, uncovering underlying causes, and examining mitigation practices.
Through an analysis of 1359 vulnerabilities from 523 System One model applications, we identify 17 vulnerability types and 34 distinct root causes across five workflow stages.
Our study further examines developers’ mitigation practices, assesses their effectiveness, and uncovers practical trade-offs between security enforcement and application functionality.
In light of these observations, we derive 12 key findings, examine their implications for securing System One model applications, and identify open research challenges in this emerging ecosystem.
\end{abstract}

%

%
\maketitle

{
\renewcommand{\thefootnote}{\fnsymbol{footnote}}
\footnotetext[2]{Corresponding author.}
}


\section{Introduction}
System One models have recently emerged as a class of AI models designed for fast, structured decision making within software applications. These models evaluate application context and return typed answers with associated probabilities, allowing developers to incorporate their predictions directly into program logic. In September 2026, TypeSafe AI introduced Jev~\cite{s1-jev-release}, the first public System One model, which supports tasks such as request routing, content classification, and workflow automation. The emerging ecosystem also includes Cloudflare’s Clef and Clef-flash~\cite{s1-clef-release} and the OpenAI Decisions API~\cite{s1-openai-release}.
These models and services enable developers to combine contextual predictions with explicit application logic, supporting automated decisions for tasks that are difficult to express through fixed rules.

However, security concerns arise when applications rely on predictions from System One models to guide their behavior. 
Shortly after Jev’s release, studies demonstrated that its decisions could be manipulated~\cite{wu2026decision,hu2026jevadvbench} and that sensitive information in application state could be inferred from its outputs~\cite{wang2026hidden}.
To support secure development, TypeSafe AI provides guidance on handling adversarial inputs in applications that use Jev~\cite{jev_jaggedness}. Its recommendations emphasize explicit decision criteria and thorough testing before deployment, with particular attention to application behavior in the presence of injected instructions or deliberately misleading content.
Despite this guidance, later reports indicate that applications using System One models still exhibit security failures, including manipulation of investment recommendations~\cite{checkpoint2026jev}, distortion of code security assessments~\cite{cirriustech2026judge}, and cross-tenant data leakage~\cite{cleffa2026security}.

Given the recent emergence of System One models, the landscape of security vulnerabilities across applications using such models remains largely unknown.
Previous studies primarily examine the security properties and capabilities of System One models, focusing on decision reliability and calibration~\cite{liu2026evaluating}, security and privacy risks~\cite{wang2026hidden}, and effectiveness in detecting AI alignment failures~\cite{guo2026just}.
However, these studies do not systematically characterize vulnerabilities in applications using System One models, their underlying causes, or the mitigation strategies adopted in practice. Understanding how these vulnerabilities arise and are addressed can help developers, maintainers, and security researchers design detection and testing tools, develop mitigation techniques, strengthen application security, and manage risks as the ecosystem evolves.

In this paper, we aim to fill these gaps by answering the following research questions:

\noindent
\textbf{RQ1:} What is the landscape of security vulnerabilities in applications using System One models?

\noindent
\textbf{RQ2:} How do security vulnerabilities arise in applications using System One models?

\noindent
\textbf{RQ3:} What mitigation strategies are adopted in applications using System One models?

To answer these questions, we systematically collect and analyze 1359 vulnerabilities from 523 GitHub repositories that integrate System One models, characterizing their security impacts and identifying 17 vulnerability types across four security functions.
We then trace these vulnerabilities through the System One model application workflow, identifying 34 distinct root causes across five workflow stages.
To understand how these vulnerabilities are addressed in practice, we examine developers’ responses, characterize six classes of mitigation strategies, and assess their effectiveness in blocking the reported exploitation paths.
We further investigate the practical challenges of vulnerability mitigation, particularly the trade-offs between security enforcement and application functionality.
Drawing on these analyses, we derive 12 key findings that reveal fundamental security challenges in System One model applications and provide insights to guide secure application design, effective vulnerability mitigation, and future security research.
\noindent
\textbf{Contributions.} We summarize the contributions as follows:
\begin{itemize}[leftmargin=12pt]
    \item We present the first systematic measurement study of security vulnerabilities in System One model applications, characterizing the vulnerability landscape and uncovering underlying failure mechanisms across the application workflow.
    \item We establish an empirical understanding of vulnerability mitigation in System One model applications, identifying recurring mitigation strategies and revealing the extent to which security risks persist despite corrective efforts.
    \item We uncover practical challenges in vulnerability remediation, highlighting the trade-offs between security enforcement and application functionality that lead developers to accept residual security risks or defer stricter safeguards.
    \item We draw on our empirical findings to identify fundamental security challenges in System One model applications and outline future research directions to advance their secure design, implementation, and analysis.
\end{itemize}

\section{Background}
\subsection{System One Model Application}

\begin{figure}[t]
    \centering
    \includegraphics[width=0.925\columnwidth]{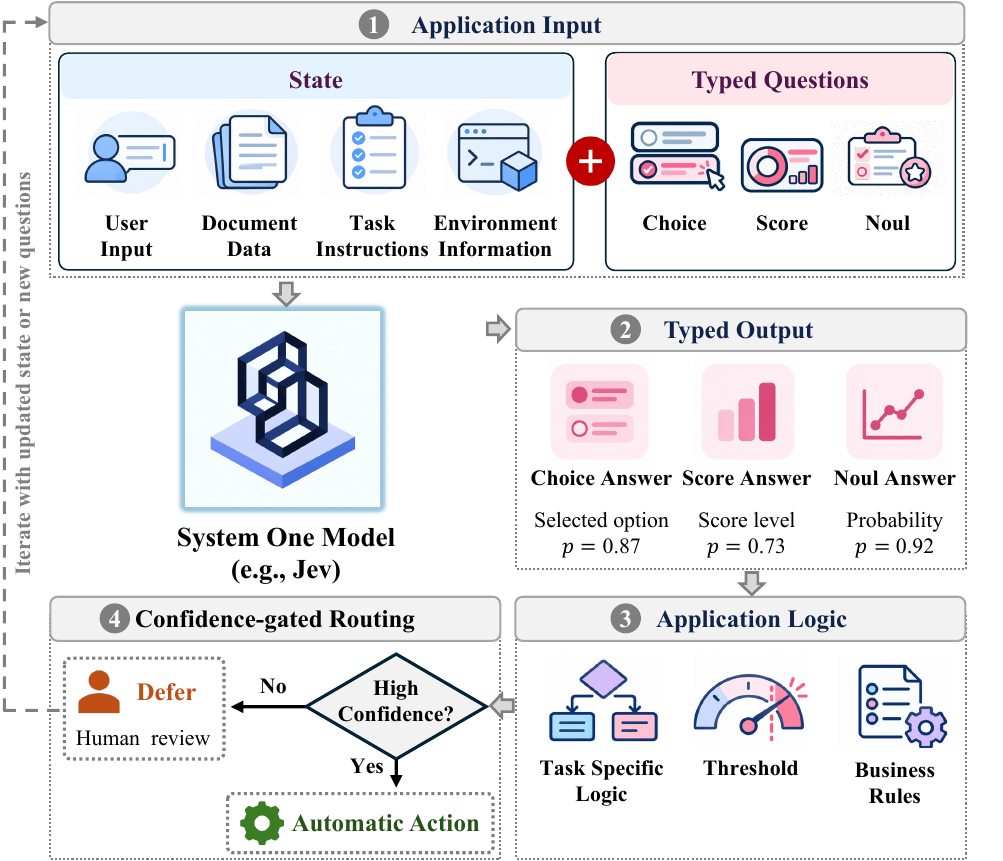}
    \vspace{-0.5em}
    \caption{\textbf{Workflow of a System One model application.}}
    \vspace{-1em}
    \label{fig:system_one_workflow}
    
\end{figure}

A System One model application incorporates models such as Jev into software workflows, where they return typed answers and probabilities rather than free-form text~\cite{s1-jev-release}.
To obtain these outputs, the application supplies a \textit{state} that contains task-relevant content and context, together with typed \textit{questions} that specify the judgments to perform.
For example, an application built on Jev expresses these judgments through three question types: \texttt{Choice} selects from predefined options, \texttt{Score} evaluates the state against an ordered rubric, and \texttt{Noul} estimates the probability that a statement is true.
The application can submit multiple questions about the same state for independent, parallel evaluation and combine the returned answers in code to determine subsequent actions.
It can further use \textit{confidence-gated routing} to decide whether to act automatically or defer a decision to human review.
The overall workflow of a System One model application is shown in Figure~\ref{fig:system_one_workflow}.

\subsection{Security Vulnerabilities}

A security vulnerability is a weakness in an application's design, implementation, or configuration that can be exploited to violate its intended security properties\bo{~\cite{zhao2026parasites, kaya2026ai, luo2025universal, 10.1145/3774904.3792409}}. 
These properties often depend on multiple software components and the interfaces between them. 
A flaw in one component may therefore compromise security guarantees enforced or assumed elsewhere in the application. Consequently, the security impact of a vulnerability depends not only on the affected component but also on its interactions with the broader application workflow.

In System One model applications, the relevant security properties span the boundary between the model and the surrounding software.
The application constructs the state and questions submitted to the model, interprets the returned judgments, and determines subsequent actions. Although the model informs these decisions, the application remains responsible for enforcing its security requirements. Vulnerabilities may arise when application logic fails to preserve the intended relationship between the evaluated state, the model judgment, and the resulting action. 
These considerations motivate a security analysis of the application as a whole rather than the model in isolation.

\section{Methodology}
\subsection{Research Questions}

The integration of System One models into applications is still at an early stage, and the types and severity of vulnerabilities in these applications remain insufficiently understood. A systematic measurement study can characterize the vulnerability landscape of such applications, clarify the security risks posed by these vulnerabilities, and identify priorities for further research. Consequently, we study the following research question:

\begin{rqbox}
\textbf{RQ1:}\enspace
\textit{What is the landscape of security vulnerabilities in
applications using System One models?}
\end{rqbox}

Addressing these security risks requires an understanding of the weaknesses that give rise to vulnerabilities and the conditions that enable exploitation. An analysis of the affected application components and exploitation mechanisms can reveal where changes to application design and implementation are needed and guide more effective vulnerability detection and mitigation. We therefore study the next research question:

\begin{rqbox}
\textbf{RQ2:}\enspace
\textit{How do security vulnerabilities arise in applications using System One models?}
\end{rqbox}

At this early stage, developers may have begun to adopt measures within their applications to mitigate these vulnerabilities. However, it remains unclear which strategies are adopted in practice, how they address the underlying weaknesses, and whether their use involves practical challenges or trade-offs. Understanding these practices can help identify approaches for mitigating similar vulnerabilities and inform future risk management. This leads to the final research question:

\begin{table*}[t]
  \centering
  \footnotesize
  \caption{Search keywords used to identify System One model applications.}
  \vspace{-1em}
  \label{tab:s1-keywords}
  \begin{tabularx}{\textwidth}{@{}
     >{\raggedright\arraybackslash}p{2.85cm}
    >{\raggedright\arraybackslash}p{2.05cm}
    >{\raggedright\arraybackslash}p{1.50cm}
    >{\raggedright\arraybackslash}X@{}}
    \toprule
    \textbf{System One model} &
    \textbf{Provider} &
    \textbf{Release date} &
    \textbf{Search keywords} \\
    \midrule

    \textbf{Jev}~\cite{s1-jev-release} &
      TypeSafe AI & 15~Sep~2026 &
      \texttt{typesafe-ai/jev}, \texttt{typesafe/jev},
      \texttt{typesafe\_sdk}, \texttt{@typesafe-ai/sdk},
      \texttt{api.typesafe.ai}, \texttt{jev-1.14} \\
    \addlinespace[1pt]

    \textbf{Kev}~\cite{s1-kev-release} &
      Jared Palmer & 17~Sep~2026 &
      \texttt{jaredpalmer/kev},
      \texttt{kev-\{0.5b,0.8b,4b,9b,27b\}} \\
    \addlinespace[1pt]

    \textbf{Bespoke-Nimble}~\cite{s1-nimble-release} &
      Bespoke Labs & 18~Sep~2026 &
      \texttt{Bespoke-Nimble},
      \texttt{bespokelabs/Bespoke-Nimble-9B},
      \texttt{bespokelabsai/nimble} \\
    \addlinespace[1pt]

    \textbf{Laya}~\cite{s1-laya-release} &
      ConvAI Innovations & 18~Sep~2026 &
      \texttt{\{convaiinnovations,nandhakishorm\}/laya},
      \texttt{LayaClient}, \texttt{laya\{.load,-serve\}} \\
    \addlinespace[1pt]

    \textbf{Decision 1.0}~\cite{s1-vllm-release} &
      vLLM & 22~Sep~2026 &
      \texttt{vllm-sr/Decision-1.0}, \texttt{vllm-sr/d3},
      \texttt{d3-flash}, \texttt{d3-mini}, \texttt{d3-nano} \\
    \addlinespace[1pt]

    \textbf{Solar Decide}~\cite{s1-solar-release} &
      Upstage & 22~Sep~2026 &
      \texttt{solar-decide}, \texttt{solar-decide-flash},
      \texttt{"upstage" "systemone"} \\
    \addlinespace[1pt]

     \textbf{Flash / Pro}~\cite{s1-telnyx-release} &
      Telnyx & 24~Sep~2026 &
      \texttt{telnyx/decision-flash},
      \texttt{telnyx/decision-pro},
      \texttt{api.telnyx.com/v2/ai/typesafe} \\
    \addlinespace[1pt]

    \textbf{Drex}~\cite{s1-drex-release} &
      Nace.AI & 24~Sep~2026 &
      \texttt{nace-ai/drex-decision-models},
      \texttt{drex-dlm}, \texttt{drex-v2.0},
      \texttt{drex-max} \\
    \addlinespace[1pt]

    \textbf{GLiNER2.5-Decide}~\cite{s1-gliner-release} &
      Fastino & 24~Sep~2026 &
      \texttt{gliner2.5-decide},
      \texttt{gliner2.5-multi-decide},
      \texttt{fastino/gliner2.5} \\
    \addlinespace[1pt]

    \textbf{Julia-1}~\cite{s1-julia-release} &
      Supersonic Labs & 25~Sep~2026 &
      \texttt{SupersonicLabs/Julia-1},
      \texttt{JuliaDecisionModel},
      \texttt{julia.load\_model} \\
    \addlinespace[1pt]

    \textbf{OneJev}~\cite{s1-onejev-release} &
      OmniJev & 27~Sep~2026 &
      \texttt{OmniJev/OneJev},
      \texttt{OneJev1.5}, \texttt{OneJev1.6} \\
    \addlinespace[1pt]

    \textbf{d1}~\cite{s1-d1-release} &
      Liquid AI & 29~Sep~2026 &
      \texttt{liquid/d1}, \texttt{liquid-ai/d1},
      \texttt{d1-ultra}, \texttt{d1.2},
      \texttt{d1.2-ultra} \\
    \addlinespace[1pt]

    \textbf{GLiDE}~\cite{s1-glide-release} &
      Fastino & 30~Sep~2026 &
      \texttt{fastino/glide}, \texttt{"Fastino" "GLiDE"} \\
    \addlinespace[1pt]

    \textbf{Mercury Decide}~\cite{s1-mercury-release} &
      Inception AI & 30~Sep~2026 &
      \texttt{mercury-decide},
      \texttt{"Inception" "Mercury Decide"} \\
    \addlinespace[1pt]

    \textbf{Trio-Spark}~\cite{s1-trio-release} &
      MachineFi Labs & 30~Sep~2026 &
      \texttt{trio-spark}, \texttt{trio-spark-preview},
      \texttt{TRIO\_SPARK\_API\_KEY},
      \texttt{/api/spark/v1/decisions} \\
    \addlinespace[1pt]

    \textbf{Clef}~\cite{s1-clef-release} &
      Cloudflare & 1~Oct~2026 &
      \texttt{@cf/cloudflare/clef},
      \texttt{cloudflare/clef}, \texttt{clef-flash},
      \texttt{clef-omni} \\
    \addlinespace[1pt]

    \textbf{Pplx-decider}~\cite{s1-pplx-release} &
      Perplexity & 1~Oct~2026 &
      \texttt{pplx-decider}, \texttt{pplx-decider-flash},
      \texttt{perplexity-ai/pplx-decider} \\
    \addlinespace[1pt]

    \textbf{Strands Decider}~\cite{s1-strands-release} &
      Strands & 1~Oct~2026 &
      \texttt{strands-decider},
      \texttt{strands-labs/strands-decider},
      \texttt{StrandsAgents/strands-decider} \\
    \addlinespace[1pt]

    \textbf{GPT-6 Luna}~\cite{s1-openai-release} &
      OpenAI & 6~Oct~2026 &
      \texttt{client.decisions.create},
      \texttt{/v1/decisions}, \texttt{"openai" "decisions"} \\
    \addlinespace[1pt]

    \textbf{Celeris-1 Decision}~\cite{s1-celeris-release} &
      Celeris & 8~Oct~2026 &
      \texttt{celeris-1-decision},
      \texttt{celeris-1.1-decision} \\
    \addlinespace[1pt]

    \textbf{Largha}~\cite{s1-largha-release} &
      Phocinae Project & 8~Oct~2026 &
      \texttt{Phocinae-Largha},
      \texttt{Phocinae/Phocinae-Largha-150M-v1},
      \texttt{phocinae-server} \\
    \addlinespace[1pt]

    \textbf{Microsoft-Decision-1}~\cite{s1-microsoft-release} &
      Microsoft & 9~Oct~2026 &
      \texttt{Microsoft-Decision-1},
      \texttt{Microsoft-Decision-1.1},
      \texttt{microsoft/microsoft-decision-1} \\
    \addlinespace[1pt]

    \textbf{SubQ Snap}~\cite{s1-subq-release} &
      Subquadratic & 12~Oct~2026 &
      \texttt{subq-snap-preview}, \texttt{subq-snap-1.1},
      \texttt{subq-snap-1.2}, \texttt{subq-snap-latest} \\
    \addlinespace[1pt]

    \textbf{Mammoth / Elephant}~\cite{s1-baseten-release} &
      Baseten & 21~Oct~2026 &
      \texttt{baselabs}, \texttt{mammoth-1},
      \texttt{elephant-1}, \texttt{bison-1},
      \texttt{groundhog-1} \\
    \addlinespace[1pt]

    \textbf{Sarvam Prism}~\cite{s1-sarvam-release} &
      Sarvam AI & 3~Nov~2026 &
      \texttt{sarvam-prism}, \texttt{"Sarvam Prism"} \\
    \addlinespace[1pt]

    \textbf{Gemini Decision 1}~\cite{s1-gemini-release} &
      Google & 11~Nov~2026 &
      \texttt{gemini-decision-1},
      \texttt{"Gemini Decisions API"},
      \texttt{"google" "decisions.create"} \\
    \addlinespace[1pt]

    \textbf{Snowflake Decision}~\cite{s1-snowflake-release} &
      Snowflake & 20~Nov~2026 &
      \texttt{snowflake-decision},
      \texttt{snowflake-decision-fast},
      \texttt{"AI\_COMPLETE" "snowflake-decision"} \\

    \bottomrule
  \end{tabularx}
\end{table*}

\begin{rqbox}
\textbf{RQ3:}\enspace
\textit{What mitigation strategies are adopted in applications using System One models?}
\end{rqbox}

\subsection{Data Collection}
\label{sec:data_collection}

\begin{figure}[t]
  \centering
  \includegraphics[width=\columnwidth]{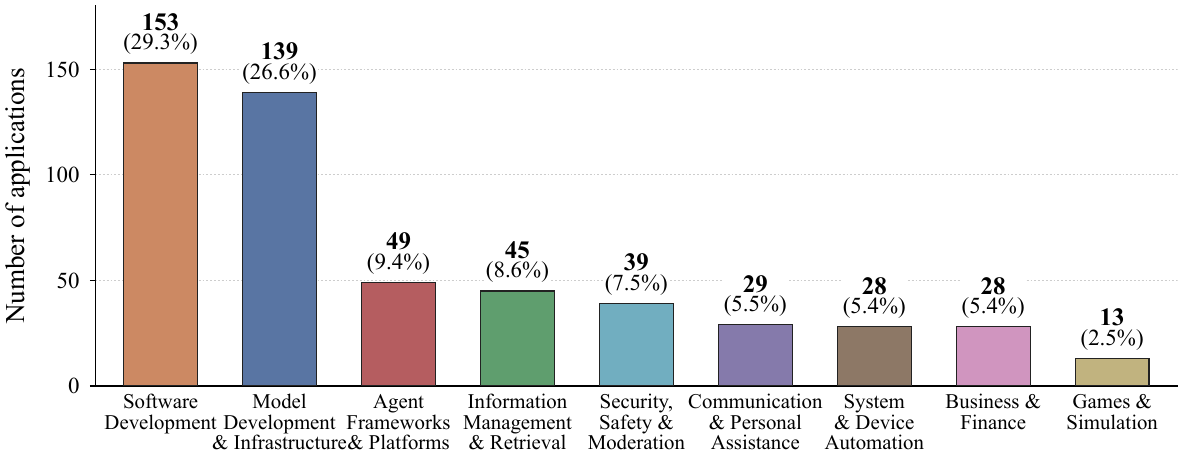}
  \vspace{-1.5em}
  \caption{Distribution of the identified applications using System One models.}
  \label{fig:fig2}
    \vspace{-2em}
\end{figure}

\noindent
\ding{182} \textbf{System One Model Application Identification.}
We search GitHub to identify applications that use System One models, with model identifiers, SDK package names, and API endpoint strings. Table~\ref{tab:s1-keywords} summarizes the complete set of keywords. The observation period was from the public release of Jev, the first System One model, on 15 September 2026 to 26 November 2026.
We then examine call signatures in the source code and excluded applications without evidence of model integration.
The final dataset comprised 523 applications, across 9 usage categories, as shown in Figure~\ref{fig:fig2}.

\noindent
\ding{183} \textbf{Vulnerability Collection.}
We collected vulnerabilities in the identified applications from two sources. First, we searched the GitHub Advisory Database~\cite{git-adv-db} for security advisories associated with these applications. Second, we searched GitHub Issues in each repository with keywords such as “security”, “vulnerability”, and "susceptibility" to capture vulnerabilities that may not be documented in formal advisories.
Note that we also searched the MITRE CVE database~\cite{mitre-cve} but found no relevant records, possibly due to the limited time for formal vulnerability disclosure at this early stage of the System One model ecosystem.



\noindent
\ding{184} \textbf{Vulnerability Screening.}
To obtain vulnerabilities relevant to System One models, we independently assessed each candidate vulnerability against two inclusion criteria. A candidate is retained only if (1) its description or associated fix provides evidence of a concrete security weakness on a code path related to System One models and (2) the affected code is merged into the repository or included in a release.
These paths cover application code for decision logic, model input construction, response handling, and credential handling, as well as model implementation, loading, and serving.
We then removed duplicates and retained 1359 distinct vulnerabilities from 523 repositories for subsequent analysis.
For each vulnerability, we record its report identifier, affected application, description, source references, available report, and remediation dates.
Moreover, we manually characterize each vulnerability in terms of its type, root cause, developer response, and mitigation strategy and effectiveness.
The criteria and procedures used for this characterization are detailed in Sections~\ref{sec:vuln-types} through \ref{sec:mitigation}.

Since most vulnerabilities collected during the initial adoption of System One models lack published CWE assignments and CVSS scores, we manually derive CWE mappings and characterize vulnerability severity.
For CWE mapping, we identify the underlying causes of the reported vulnerabilities and select the most specific applicable CWE entries in accordance with MITRE’s root cause mapping guidance~\cite{mitre-guidance}.
A vulnerability’s CWE mapping remains unspecified when information on its root cause is insufficient to support a reliable assignment.
For severity characterization, we assess the consequences of successful exploitation for the confidentiality, integrity, and availability of the vulnerable application according to the CVSS v4.0 impact criteria~\cite{cvss-doc}. Each dimension is rated independently as None, Low, or High based on the consequences supported by the available evidence. An impact dimension remains undetermined when the evidence is insufficient to support a reliable rating. 
Note that we report application impact ratings rather than CVSS scores as the available reports do not consistently document the exploitation conditions required for reliable scoring.

\begin{table}[t]
\centering
\caption{Distribution of vulnerability types in System One model applications.}
\vspace{-1em}
\label{tab:vulnerability_dist}
\small
\renewcommand{\arraystretch}{1}
\begin{tabularx}{\columnwidth}{
    >{\raggedright\arraybackslash}X
    >{\raggedleft\arraybackslash}p{1cm}@{}
}
\toprule
\textbf{Vulnerability Type} & \textbf{Count} \\
\midrule
V1: Fail-Open Decision Logic & 78 \\
V2: Incomplete Policy Enforcement & 172 \\
V3: Improper Model Response Validation & 40 \\
V4: Incorrect Judgment Criteria and Interpretation & 29 \\
V5: Improper Judgment Context Construction & 70 \\
V6: Security Configuration Integrity Failures & 69 \\
V7: Prompt and Evidence Trust Boundary Violations & 35 \\
V8: Improper Decision Binding and Reuse & 45 \\
V9: Unintended Sensitive Data Transmission & 202 \\
V10: Private Application State Exposure & 174 \\
V11: Inadequate Provider Credential and Channel Protection & 76 \\
V12: Improper Service Authentication and Origin Validation & 152 \\
V13: Inadequate Resource and Availability Controls & 147 \\
V14: Unsafe Execution and Output Rendering & 27 \\
V15: Improper Filesystem Confinement and Path Validation & 29 \\
V16: Memory Safety Violations & 9 \\
V17: Insufficient Security Audit Logging & 5 \\
\midrule
\textbf{Total} & \textbf{1359} \\
\bottomrule
\end{tabularx}
\vspace{-1em}
\end{table}

\subsection{Vulnerability Types}
\label{sec:vuln-types}

As described in Section~\ref{sec:data_collection}, we identified 135 distinct CWEs by analyzing the underlying weaknesses of the collected vulnerabilities.
However, the direct use of these CWEs as vulnerability types poses several issues.
First, some CWE entries lack sufficient granularity to distinguish application-level security failures that arise from different underlying mechanisms.
For example, CWE-863 \textit{Incorrect Authorization} can describe both an overbroad allowlist that bypasses required model-based checks and application logic that substitutes a model verdict for mandatory human approval. The former omits a required security check, whereas the latter grants the model approval authority reserved for a human reviewer. A classification based solely on this identifier would therefore obscure these distinct failure mechanisms.

Second, closely related application-level weaknesses map to different CWEs. For instance, failure to verify the expected type of a model answer corresponds to CWE-1287 \textit{Improper Validation of Specified Type of Input}, whereas failure to check that choice probabilities sum to one corresponds to CWE-1288 \textit{Improper Validation of Consistency within Input}. Although these failures concern different response properties, both reflect inadequate validation of model responses and therefore warrant a common vulnerability type.

Third, certain CWEs exhibit conceptual overlap due to differences in granularity.
For example, CWE-313 \textit{Cleartext Storage in a File or on Disk} is a specialization of CWE-312 \textit{Cleartext Storage of Sensitive Information}. These CWEs therefore do not define mutually exclusive vulnerability types, as the former describes a specific form of the weakness covered by the latter.
Finally, there are vulnerabilities for which the root cause information is insufficient to support a reliable CWE mapping.

Hence, we classify the collected vulnerabilities based on their underlying failure mechanisms and affected security functions, inspired by previous works on software vulnerability classification~\cite{bo2025understanding,yang2026poisoned,meisami2025sigscope}. We consolidate related defects within a common security function and preserve distinctions between failures that violate different security requirements. Each vulnerability receives a primary type according to its principal security defect, while available CWE mappings serve as supplementary annotations. 
For aggregate analysis, we further group the resulting 17 vulnerability types into four security functions according to the application security responsibility they primarily affect, drawing on the security control families defined in NIST SP 800-53~\cite{nist80053r5}.
Table~\ref{tab:vulnerability_dist} summarizes the resulting vulnerability types and their frequencies.

\begin{table*}[t]
\centering
\caption{Root causes of vulnerabilities in System One model applications.}
\vspace{-1em}
\label{tab:root_causes}
\small
\begin{tabularx}{\textwidth}{
    >{\raggedright\arraybackslash}p{0.32\textwidth}
    >{\raggedright\arraybackslash}X
    >{\raggedleft\arraybackslash}p{0.065\textwidth}@{}
}
\toprule
\textbf{Workflow Stage} &
\textbf{Failure Mechanism} &
\textbf{Count} \\
\midrule

\multirow[t]{9}{=}{\raggedright
    \textbf{R1: Judgment Preparation}
    \par (Total: 379)
}
& \textbf{R1-1:} Incomplete or inaccurate judgment context & 68 \\
& \textbf{R1-2:} Untrusted content treated as instructions, intent, or evidence & 33 \\
& \textbf{R1-3:} Inadequate configuration validation and protection & 82 \\
& \textbf{R1-4:} Insufficient model input screening and redaction & 154 \\
& \textbf{R1-5:} Incomplete or misaligned judgment questions and criteria & 6 \\
& \textbf{R1-6:} Private state exposure via credential setup and storage & 19 \\
& \textbf{R1-7:} Insufficient namespace confinement during file and helper preparation & 9 \\
& \textbf{R1-8:} Unenforced model artifact integrity and format requirements & 5 \\
& \textbf{R1-9:} Unsafe deserialization of model artifacts & 3 \\

\midrule

\multirow[t]{7}{=}{\raggedright
    \textbf{R2: Judgment Acquisition and Validation}
    \par (Total: 402)
}
& \textbf{R2-1:} Incomplete model interface data validation & 104 \\
& \textbf{R2-2:} Insufficient caller, origin, or usage authorization in judgment services & 63 \\
& \textbf{R2-3:} Violations of credential or recipient constraints in provider requests & 141 \\
& \textbf{R2-4:} Private information exposure through provider request or error handling & 38 \\
& \textbf{R2-5:} Insufficient resource admission, accounting, or failure containment & 43 \\
& \textbf{R2-6:} Native buffer bounds violation during inference request & 9 \\
& \textbf{R2-7:} Insufficient isolation of mutable private state in judgment clients & 4 \\

\midrule

\multirow[t]{5}{=}{\raggedright
    \textbf{R3: Decision Composition}
    \par (Total: 121)
}
& \textbf{R3-1:} Incorrect judgment interpretation or aggregation & 21 \\
& \textbf{R3-2:} Overreliance on model judgments as the sole decision basis & 4 \\
& \textbf{R3-3:} Incomplete or inconsistent deterministic policy checks & 89 \\
& \textbf{R3-4:} Unbounded policy computation & 6 \\
& \textbf{R3-5:} Executable input accepted in decision expressions & 1 \\

\midrule

\multirow[t]{3}{=}{\raggedright
    \textbf{R4: Decision Routing and Escalation}
    \par (Total: 84)
}
& \textbf{R4-1:} Unsafe continuation despite judgment errors or unavailability & 77 \\
& \textbf{R4-2:} Incorrect uncertainty routing or escalation disposition & 3 \\
& \textbf{R4-3:} Missing origin or identity checks in review disposition interfaces & 4 \\

\midrule

\multirow[t]{10}{=}{\raggedright
    \textbf{R5: Decision Enactment and Reuse}
    \par (Total: 373)
}
& \textbf{R5-1:} Required decisions omitted or misapplied at execution & 84 \\
& \textbf{R5-2:} Insufficient binding and validation of reused approvals & 49 \\
& \textbf{R5-3:} Fresh favorable judgments obtained through repeated identical requests & 1 \\
& \textbf{R5-4:} Private information exposure during execution or via retained state & 44 \\
& \textbf{R5-5:} Missing access or origin checks in action and output services & 73 \\
& \textbf{R5-6:} Unsafe handling of untrusted output at execution or rendering sinks & 23 \\
& \textbf{R5-7:} Insufficient filesystem confinement of execution and output paths & 18 \\
& \textbf{R5-8:} Insufficient sanitization of sensitive outputs, logs, or exports & 66 \\
& \textbf{R5-9:} Missing or truncated execution evidence & 5 \\
& \textbf{R5-10:} Insufficient confidentiality isolation of reused state across callers & 10 \\

\bottomrule
\end{tabularx}
\vspace{-1em}
\end{table*}

\subsection{Root Causes}
\label{sec:root-causes}

To characterize the root causes of the collected vulnerabilities, we organize our analysis at two levels.
At the workflow level, we assign each vulnerability to the stage responsible for the defective operation.
This attribution distinguishes a defect's origin from the stages in which its consequences manifest.
For example, failure to reject a malformed model response is attributed to judgment acquisition and validation, even when the invalid response subsequently leads to an incorrect decision during decision composition.

At the mechanism level, we identify the specific errors or missing safeguards responsible for each vulnerability.
These underlying causes serve as the basis for distinguishing failure mechanisms within each workflow stage.
This distinction is crucial as failures within the same workflow stage may produce similar security consequences yet arise from distinct underlying defects.
Table~\ref{tab:root_causes} summarizes the 34 distinct root causes across the 5 workflow stages.

\subsection{Mitigation}
\label{sec:mitigation}

For each collected vulnerability, we examine developers’ responses, characterize the mitigation strategies adopted, and assess their effectiveness in addressing the underlying weaknesses.
The response classification is based on developers’ engagement with vulnerability reports and their acknowledgment of the associated security risks.
Mitigation strategies are categorized according to their underlying mechanisms, and their effectiveness is evaluated based on whether the vulnerabilities remain exploitable after mitigation.

To characterize developer responses, we examine the available discussions and development records for each vulnerability.
A developer may leave a report unanswered, respond without undertaking corrective work, or initiate mitigation. 
Thus, such responses can be categorized as follows:
\begin{itemize}[leftmargin=*]
    \item \textbf{No response.} The available report history contains no developer response to the reported vulnerability.

    \item \textbf{Response only.} Developers engage with the report, but the record contains no evidence of concrete corrective work.

    \item \textbf{Mitigation initiated.} Developers undertake concrete corrective work to address the reported vulnerability. This category includes measures under development and those already applied.
\end{itemize}

For vulnerabilities in which mitigation has been initiated, we examine the mechanisms of the associated corrective measures.
Mitigation strategies are not mutually exclusive, as developers may apply different types of measures to the same vulnerability. 
We categorize them as follows:

\begin{itemize}[leftmargin=*]
\item \textbf{Failure handling.} Measures govern application behavior when model calls fail or return missing, malformed, or incomplete decisions through explicit error handling, conservative defaults, or escalation to human review.

\item \textbf{Decision constraints.} Measures introduce or revise the rules governing model inputs, outputs, and downstream actions through allowlists, mandatory policy conditions, threshold bounds, or revised judgment criteria.

\item \textbf{Data and credential hygiene.} Measures limit sensitive data and credential exposure through filtering, redaction, restrictions on storage and propagation, and checks on transmission destinations and transport security.

\item \textbf{Access control and resource limits.} Measures restrict access to application functions and security settings through authentication and permission checks, or bound resource consumption through request, rate, and spending limits.

\item \textbf{Integration fixes and observability.} Measures correct errors in model invocation, decision enforcement, and configuration, or introduce diagnostics that expose missing or failing checks.

\item \textbf{Security notices and feature disablement.} Measures communicate security risks and usage limitations through documentation and notices, or disable or remove the affected functionality.
\end{itemize}

\begin{figure}[!t]
    \centering
    \begin{minipage}[t]{0.49\columnwidth}
        \centering
        \includegraphics[width=\linewidth]
            {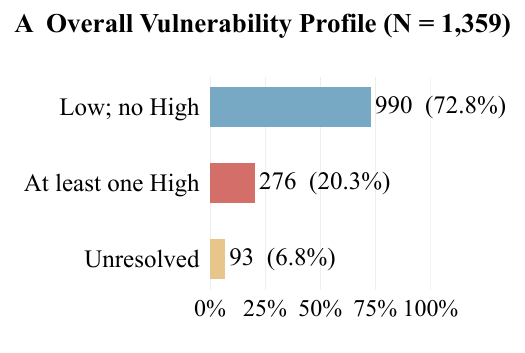}
    \end{minipage}%
    \begin{minipage}[t]{0.49\columnwidth}
        \centering
        \includegraphics[width=\linewidth]
            {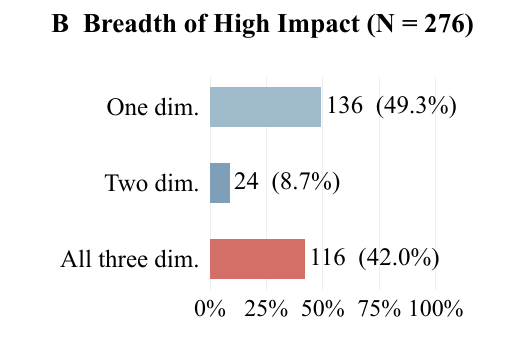}
    \end{minipage}

    \vspace{-2mm}

    \caption{
    Severity distribution and breadth of security impacts across 1,359 collected vulnerabilities.
    }
    \label{fig:security-impact}
     \vspace{-2em}
\end{figure}

Finally, we assess whether the applied mitigations prevent exploitation of the reported vulnerabilities. The assessment draws on vulnerability reports, associated code changes, and follow-up discussions to identify residual exploitation paths and configuration dependencies. 
We categorize mitigation effectiveness as follows:

\begin{itemize}[leftmargin=*]
\item \textbf{Exploitation blocked.} The available evidence indicates that the applied measures block the reported exploitation paths under the assessed conditions.

\item \textbf{Residual exploitability.} The reported vulnerability remains exploitable despite the applied corrective measures.

\item \textbf{Configuration-dependent exploitability.} The reported vulnerability remains exploitable under configurations that disable or weaken the applied security controls.
\end{itemize}

\section{Measurement Findings}

\subsection{RQ1: Security Vulnerability Landscape}
\label{sec:rq1}

We first assess the security impact of the collected vulnerabilities on the confidentiality, integrity, and availability of System One model applications. 
As shown in Figure~\ref{fig:security-impact}, 990 vulnerabilities (72.8\%) have at least one \textit{Low} but no \textit{High} impact, whereas 276 (20.3\%) have \textit{High} impact in at least one dimension. Among these 276 vulnerabilities, 140 (50.7\%) have \textit{High} impact in at least two dimensions, including 116 (42.0\%) across all three.
These results suggest that a majority of Low-impact vulnerabilities does not necessarily translate into acceptable security risks for an intended deployment. 
For organizations considering System One model applications for security-sensitive deployments, the potential for High-impact vulnerabilities to compromise multiple security properties may pose a barrier to adoption, despite their limited prevalence.

\begin{table*}[t]
\centering
\caption{Distribution of collected vulnerabilities across security functions.}
\vspace{-1em}
\label{tab:security-function}
\footnotesize
\begin{tabularx}{\textwidth}{@{}
    >{\raggedright\arraybackslash}p{0.22\textwidth}
    >{\raggedright\arraybackslash}X
    >{\centering\arraybackslash}p{0.09\textwidth}
    >{\raggedleft\arraybackslash}p{0.07\textwidth}
    >{\raggedleft\arraybackslash}p{0.07\textwidth}@{}}
\toprule
\textbf{Security function}
    & \textbf{Security responsibility}
    & \textbf{Types}
    & \textbf{Count}
    & \textbf{Share} \\
\midrule

Decision processing and enforcement
    & Ensures the integrity of model-mediated decisions
      across context construction, response validation,
      and downstream enforcement.
    & V1--V8
    & 538
    & 39.6\% \\

Data protection and access control
    & Protects sensitive application state and credentials
      from unauthorized disclosure and restricts access
      to model providers and application services.
    & V9--V12
    & 604
    & 44.4\% \\

Runtime and resource protection
    & Constrains execution, filesystem and memory operations,
      and resource consumption to preserve safe
      and available operation.
    & V13--V16
    & 212
    & 15.6\% \\

Security logging and auditing
    & Records security-relevant events to support
      auditing and incident investigation.
    & V17
    & 5
    & 0.4\% \\

\bottomrule
\end{tabularx}
\end{table*}

\finding{1}{Although vulnerabilities with \textit{High} impact constitute a minority of the collected vulnerabilities, their consequences typically span multiple security dimensions. These vulnerabilities warrant particular attention because a single weakness can expose an application to several forms of harm.}

Next, we examine the prevalence of individual vulnerability types.
As shown in Table~\ref{tab:vulnerability_dist}, V9 \textit{Unintended Sensitive Data Transmission}, V10 \textit{Private Application State Exposure}, and V2 \textit{Incomplete Policy Enforcement} are the three most prevalent types.
System One model applications construct judgment requests from user inputs, contextual evidence, and internal application state. Overly broad context selection or inadequate restrictions on request destinations can cause unintended transmission of sensitive information (V9). 
Private application state may also appear in outputs and retained artifacts, such as logs and caches. Inadequate isolation or sanitization can expose this state to unintended parties (V10).
Moreover, a returned model judgment does not itself enforce application policy. Missing policy checks or unguarded execution paths can allow actions to proceed without the intended restrictions (V2).
These patterns reveal a recurring gap between intended security restrictions and their enforcement across sensitive information flows and downstream actions.

\finding{2}{\textit{Unintended Sensitive Data Transmission}, \textit{Private Application State Exposure}, and \textit{Incomplete Policy Enforcement} are the three most prevalent vulnerability types in System One model applications. They reflect insufficient control over transmitted information, inadequate protection of private state in outputs and retained artifacts, and incomplete enforcement of policy across execution paths.}

\begin{figure}[!t]
\centering
\includegraphics[width=\columnwidth]
    {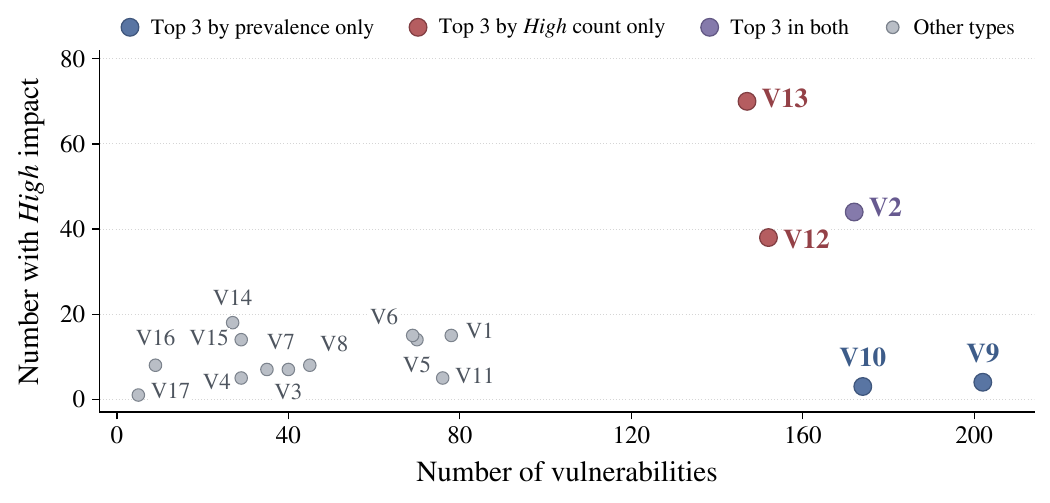}
    \vspace{-2em}
    \caption{Relationship between vulnerability security impact and type prevalence.}
    \label{fig:type-prevalence-impact}
    \vspace{-1.5em}
\end{figure}

We also examine the relationship between vulnerability type prevalence and security impact.
As shown in Figure~\ref{fig:type-prevalence-impact}, V13 \textit{Inadequate Resource and Availability Controls}, V2 \textit{Incomplete Policy Enforcement}, and V12 \textit{ Improper Service Authentication and Origin Validation} account for the largest numbers of vulnerabilities rated High in at least one security dimension, with 70, 44, and 38, respectively. Although these three types represent 471 of the 1359 vulnerabilities (34.7\%), they account for 152 of the 276 \textit{High}-impact vulnerabilities (55.1\%). 
Among these 276 vulnerabilities, 116 (42.0\%) are rated High across all three security dimensions.
Conversely, V9 and V10, the two most prevalent vulnerability types, account for only seven \textit{High}-impact vulnerabilities (2.5\%). 
Notably, V2 is the only vulnerability type ranked among the top three in both prevalence and the number of \textit{High}-impact vulnerabilities, highlighting the substantial security risks associated with insufficient control over downstream actions.

\finding{3}{Vulnerability type prevalence does not fully capture the distribution of High impact. \textit{Incomplete Policy Enforcement} contributes disproportionately to \textit{High}-impact vulnerabilities, revealing the severe security risks of treating model judgments as sufficient authorization for downstream actions.}

We further examine how the collected vulnerabilities are distributed across the security functions they primarily affect. 
As shown in
Table~\ref{tab:security-function}, 604 of the 1359 vulnerabilities (44.4\%) affect data protection and access control, and 538 (39.6\%) affect decision processing and enforcement. In contrast, 212 vulnerabilities (15.6\%) affect runtime and resource protection, and only five vulnerabilities (0.4\%) affect security logging and auditing. 
Despite the diversity of underlying weaknesses, this distribution reveals that vulnerabilities are concentrated in two core responsibilities of System One model applications. Most of them affect either the protection of sensitive information and control over service access or the integrity of decisions informed by model judgments.

\finding{4}{The vulnerability landscape is diverse in mechanism but concentrated in security function. Most vulnerabilities affect the controls governing model judgment acquisition and downstream enforcement, highlighting model integration as a central security concern.}




\subsection{RQ2: Root Causes}
\label{sec:rq2}

Having characterized the vulnerability landscape, we now examine the underlying causes of the collected vulnerabilities across the application workflow.
As shown in Table~\ref{tab:root_causes}, R1 \textit{Judgment Preparation} and R2 \textit{Judgment Acquisition and Validation} account for 379 (27.9\%) and 402 (29.6\%) vulnerabilities, respectively. 
Both stages precede R3 \textit{Decision Composition}, with the former governing the evidence supplied to the model and the latter the integrity of model interactions. 
This concentration indicates that a substantial proportion of vulnerabilities originate in the preparation and acquisition of model judgments, before applications use them to make decisions or guide subsequent actions.

Within R1 \textit{Judgment Preparation}, R1-4 \textit{insufficient model input screening and redaction} is the most prevalent mechanism, accounting for 154 of the 379 vulnerabilities in this stage and 11.3\% of the total.
System One model applications construct judgment context from user inputs, files, configuration data, and internal application state. 
Excessive context inclusion or inadequate redaction can expose sensitive information to model providers before a judgment is returned. 
Context construction can also omit relevant evidence or misrepresent the operation being assessed. These deficiencies fall under R1-1 \textit{incomplete or inaccurate judgment context}, which accounts for another 68 vulnerabilities.
Such omissions and inaccuracies can compromise subsequent application decisions even if the model correctly evaluates the supplied context.

R2 \textit{Judgment Acquisition and Validation} encompasses a complementary set of deficiencies in model integration.
R2-3 \textit{violations of credential or recipient constraints in provider requests} (141), R2-1 \textit{incomplete model interface data validation} (104), and R2-2 \textit{insufficient caller, origin, or usage authorization in judgment services} (63) account for 76.6\% of the vulnerabilities in this stage.
The model interface therefore constitutes a security boundary, where weaknesses in credential handling, recipient restrictions, data validation, or service authorization can compromise application security.


\begin{figure}[t]
    \centering
    \begin{subfigure}[t]{0.465\columnwidth}
        \centering
        \includegraphics[width=\linewidth]{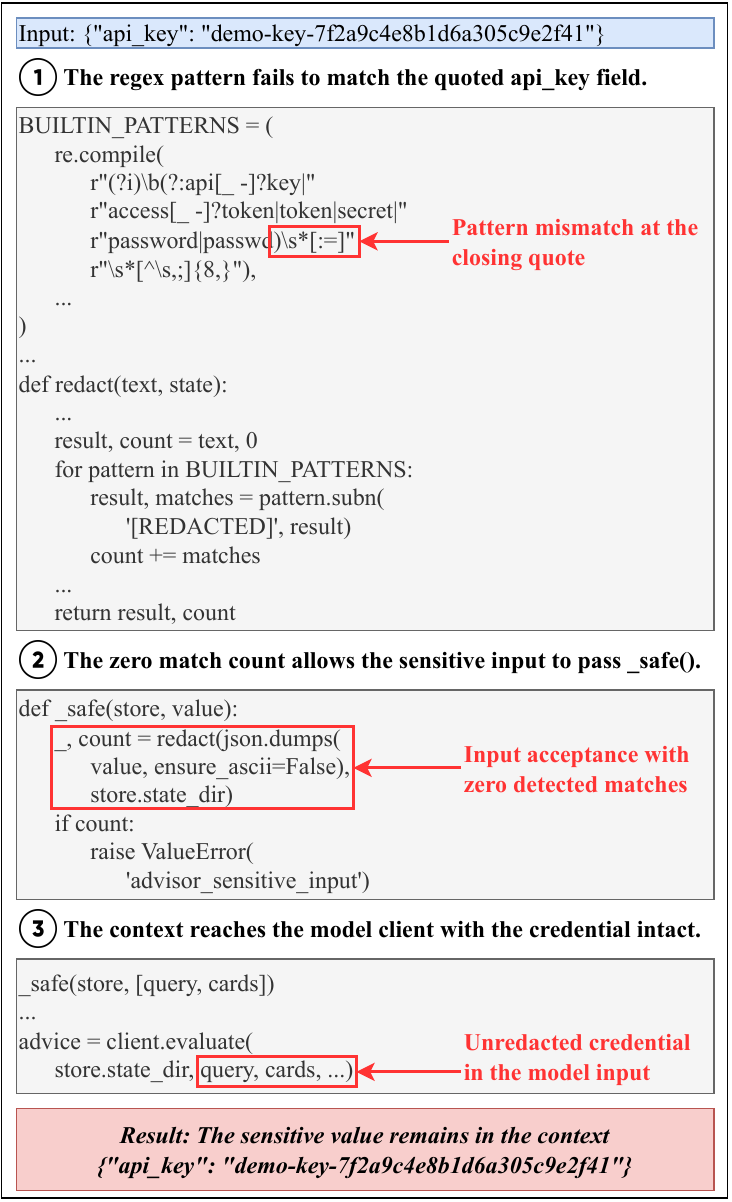}
        \caption{R1-4 Insufficient model input screening in \textsc{avenoxbeyin} pull request\#76.}
        \label{fig:r1-4}
    \end{subfigure}
    \hfill
    \begin{subfigure}[t]{0.507\columnwidth}
        \centering
        \includegraphics[width=\linewidth]{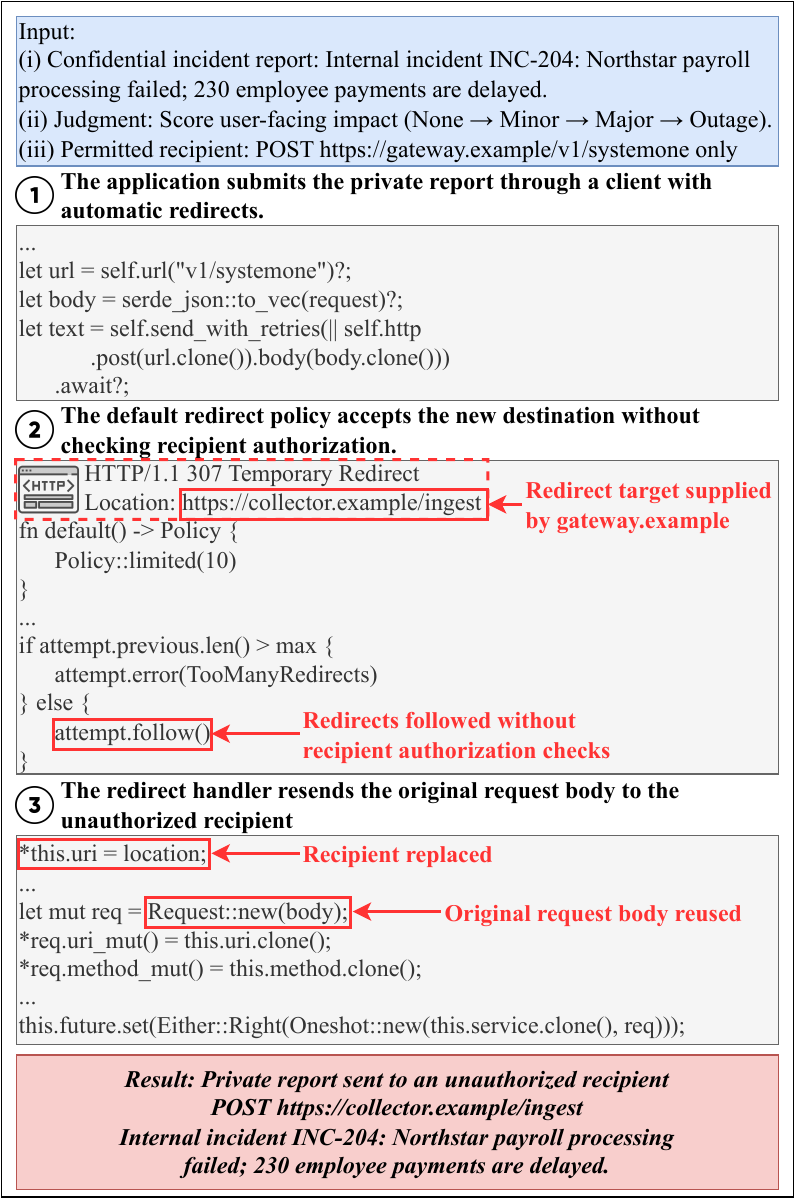}
        \caption{R2-3 Unenforced recipient constraints in \textsc{Signalman} pull request\#24.}
        \label{fig:r2-3}
    \end{subfigure}
    \caption{Root causes of sensitive information exposure during judgment preparation and acquisition.}
    \label{fig:root-causes_r1_r2}
     \vspace{-2em}
\end{figure}

To illustrate the dominant mechanisms in R1 and R2, we present two representative examples in Figure~\ref{fig:root-causes_r1_r2}.
In the \textsc{avenoxbeyin}\footnote{https://github.com/avenoxai/avenoxbeyin} application, it screens serialized judgment context for sensitive information before invoking the model client.
The screening pattern does not accommodate the closing quotation mark after \texttt{api\_key}, so the illustrated input yields zero matches. 
Since \texttt{\_safe()} rejects inputs only when a match is detected, the credential passes screening and reaches the model client intact. 
In the \textsc{signalman}\footnote{https://github.com/chussenot/signalman} application, the default redirect policy limits the number of redirects but does not enforce recipient constraints. 
An HTTP 307 response from the permitted gateway "\texttt{https://gateway.example/v1/systemone}" can therefore redirect the request to the unauthorized endpoint "\texttt{https://collector.example/ingest}" specified in the \texttt{Location} header. The application preserves the original POST body and forwards the confidential report to that endpoint.
These examples demonstrate that secure model integration requires safeguards not only for returned judgments but also for the preparation and delivery of model requests.

\finding{5}{In System One model applications, judgment context construction and model provider invocation establish critical security boundaries. Failures in input screening, credential and recipient validation, response verification, or service authorization can expose sensitive information or compromise downstream decisions, even when the model correctly evaluates the supplied context.}

Beyond deficiencies at the model boundary, R5 \textit{Decision Enactment and Reuse} accounts for the next largest concentration of root causes, with 373 vulnerabilities (27.4\%).
At this stage, security depends on application logic that enforces decisions during execution and preserves their intended scope during reuse. 
Specifically, R5-1 \textit{required decisions omitted or misapplied at execution} accounts for 84 vulnerabilities, while 49 involve R5-2 \textit{insufficient binding and validation of reused approvals}.
These defects arise when execution logic leaves some action paths outside the required decision checks or permits approval reuse without revalidation after the operation or execution context changes.
Application logic also governs access to services and the disclosure of sensitive information during execution.
Deficiencies in these controls account for 73 vulnerabilities under R5-5 \textit{missing access or origin checks in action and output services}, and another 66 under R5-8 \textit{insufficient sanitization of sensitive outputs, logs, or exports}. 
Correct judgments alone therefore do not ensure secure execution, as application logic governs the enforcement of decision constraints, control over service access, and the protection of sensitive data during execution and reuse.

\begin{figure}[t]
  \centering
  \includegraphics[width=\columnwidth]{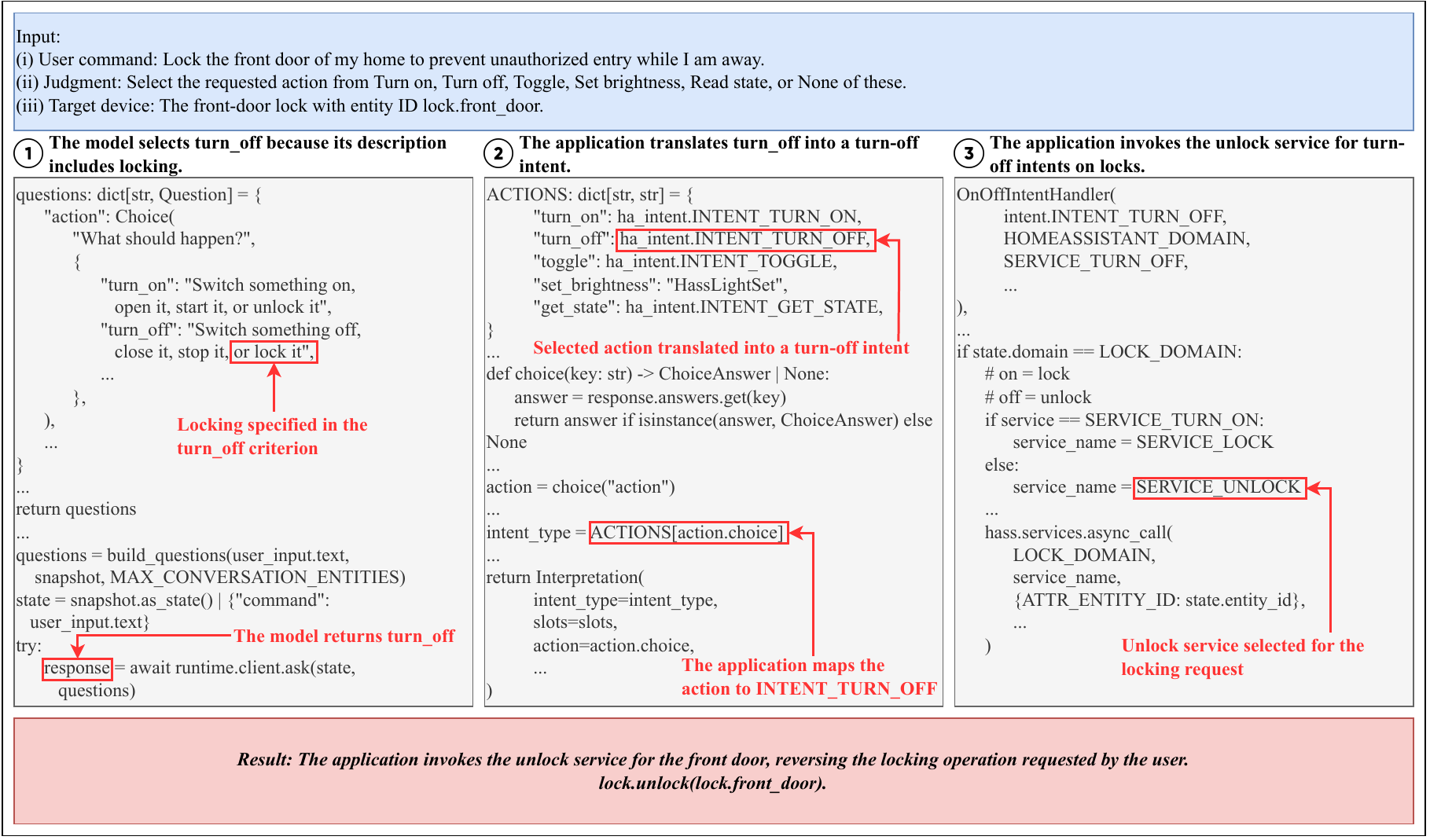}
\caption{Root cause of R5-1 decision misapplication at execution in \textsc{ha-jev} pull request\#6.}
  \label{fig:root-causes_r5}
   \vspace{-2em}
\end{figure}

Figure~~\ref{fig:root-causes_r5} presents a representative example of R5-1 in the \textsc{ha-jev}\footnote{https://github.com/AboveColin/HA-Jev} application. 
The application first asks the model which predefined action matches the user’s request to lock the front door. 
The model returns \texttt{turn\_off} because that option’s description includes “or lock it.” 
Then, the application translates the returned choice into \texttt{INTENT\_TURN\_OFF} based on \texttt{ACTIONS} dictionary. 
For lock entities, however, the application maps turn-on intents to locking and turn-off intents to unlocking, contrary to the action descriptions supplied to the model.
The application therefore selects \texttt{SERVICE\_UNLOCK} for the returned \texttt{turn\_off} choice and invokes \texttt{lock.unlock} on \texttt{lock.front\_door}, reversing the requested locking operation.
This example demonstrates that secure decision enactment requires applications to preserve the meaning of model judgments when translating them into operations, maintaining consistency between the supplied decision criteria and the resulting behavior.

\finding{6}{System One model applications depend on application logic to preserve the intended meaning and scope of judgments and enforce their constraints during execution and reuse. Failures in decision enactment can nullify the protection of otherwise valid judgments, while insecure handling of associated state and outputs can expose sensitive information.}


Finally, R3 \textit{Decision Composition} and R4 \textit{Decision Routing and Escalation} account for 121 (8.9\%) and 84 (6.2\%) vulnerabilities, respectively.
Within these stages, the dominant root causes are R3-3 \textit{incomplete or inconsistent deterministic policy checks}, with 89 cases (73.6\% of R3), and R4-1 \textit{unsafe continuation despite judgment errors or unavailability}, with 77 cases (91.7\% of R4).
These mechanisms reveal two distinct failures in application control logic.
Incomplete policy checks can extend approval from a permitted operation to effects outside the scope of those checks, while permissive error handling can bypass authorization requirements that should remain enforced when a model judgment is unavailable.

\begin{figure}[t]
    \centering
    \begin{subfigure}[t]{0.465\columnwidth}
        \centering
        \includegraphics[width=\linewidth]{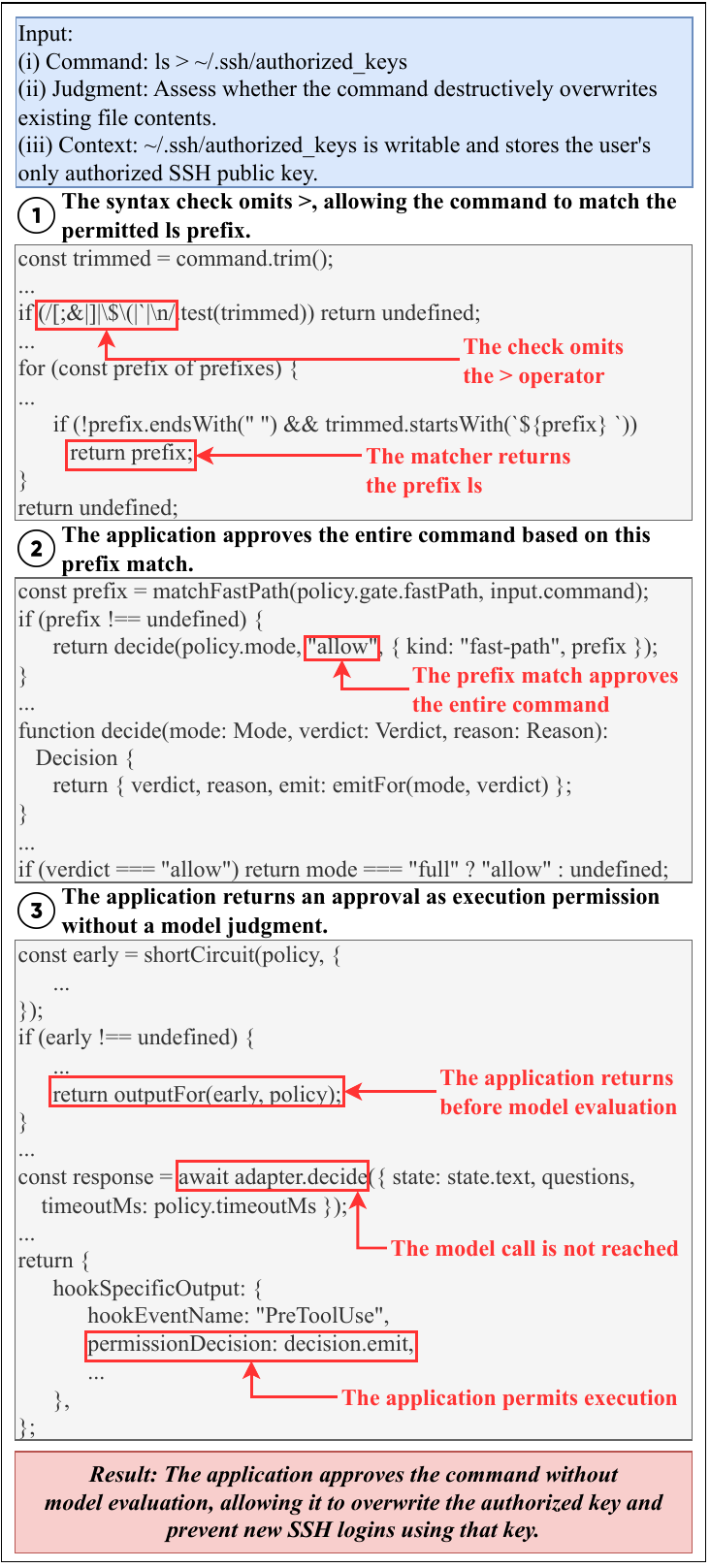}
        \caption{R3-3 Incomplete deterministic policy checks in \textsc{bouncer} pull request\#34.}
        \label{fig:r1-4}
    \end{subfigure}
    \hfill
    \begin{subfigure}[t]{0.507\columnwidth}
        \centering
        \includegraphics[width=\linewidth]{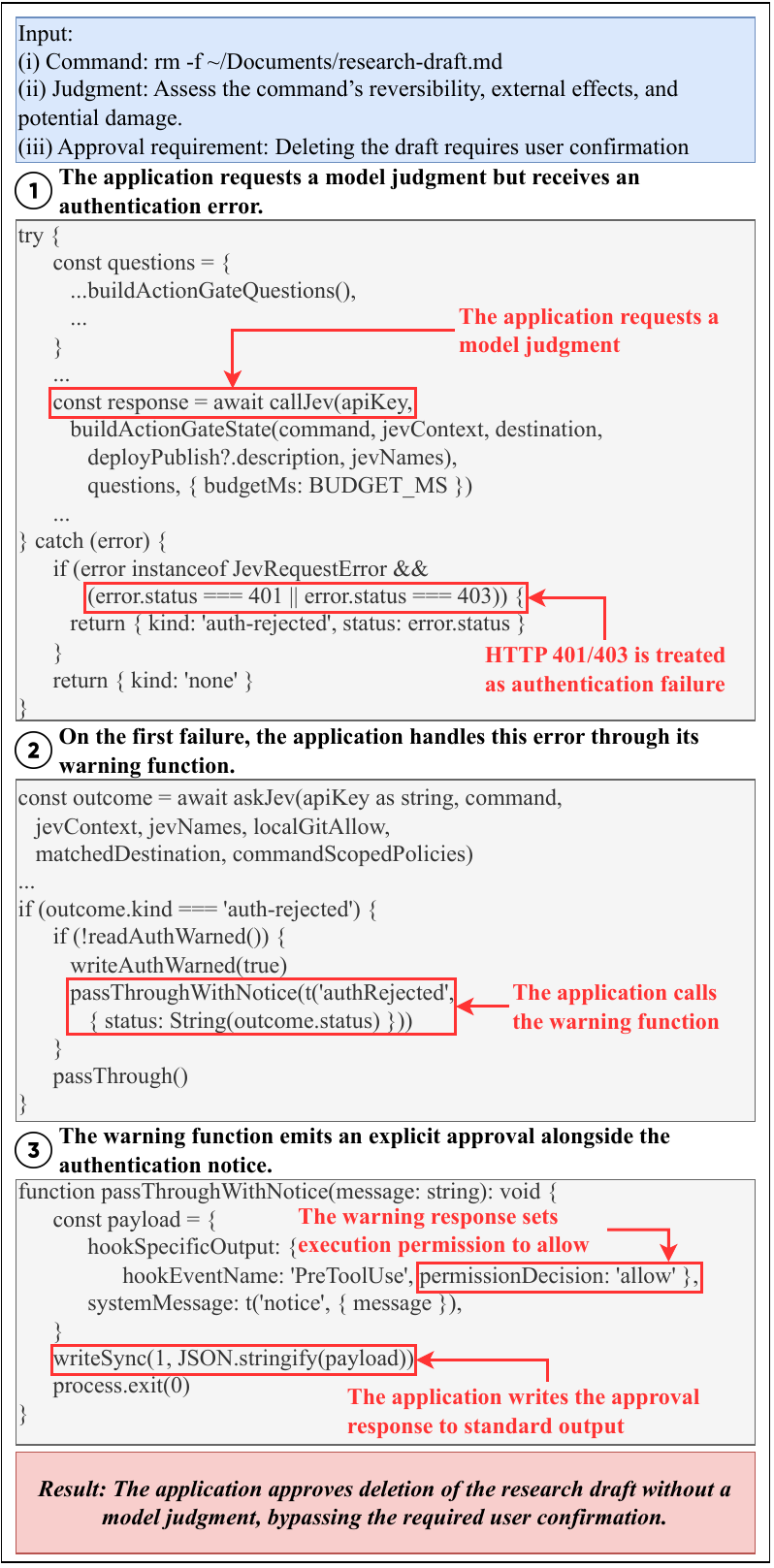}
        \caption{R4-1 Command approval despite judgment failure in \textsc{orca-jev-advisor} pull request\#28.}
        \label{fig:r2-3}
    \end{subfigure}
    \caption{Root causes of command authorization bypass during decision composition and routing.}
    \label{fig:root-causes_r3_r4}
    \vspace{-1em}
\end{figure}

Figure~\ref{fig:root-causes_r3_r4} illustrates R3-3 and R4-1 through representative examples of command authorization bypass.
In the \textsc{bouncer}\footnote{https://github.com/clownware/bouncer} application, the syntax check omits the \texttt{\textgreater{}} redirection operator, so \texttt{ls \textgreater{} \textasciitilde{}/.ssh/authorized\_keys} matches the permitted \texttt{ls} prefix. Based on this match, the application approves the entire command and returns before model evaluation. Execution of the approved command would overwrite the authorized key with a directory listing and prevent new SSH logins that rely on that key.
In the \textsc{orca-jev-advisor}\footnote{https://github.com/ab2webco/orca-jev-advisor} application, the model provider rejects the judgment request with an HTTP 401 or 403 response and returns no model judgment. The application therefore invokes a warning function that also emits \texttt{permissionDecision: 'allow'} alongside the error notice. The resulting approval permits deletion of the research draft without the required user confirmation.
Such examples show that secure decision composition and routing require applications to bind execution permission to the policy requirements of the complete operation and preserve those requirements even when model evaluation is bypassed or fails.

\finding{7}{Security controls surrounding model judgments rely on deterministic policy enforcement and safe handling of evaluation failures. When policy checks are incomplete or error handlers trigger fail-open behavior, protected operations may proceed without valid authorization, even in the absence of incorrect model judgments.}

\subsection{Mitigation}

We first examine developers’ responses to reported vulnerabilities and their subsequent mitigation efforts. As shown in Table~\ref{tab:mitigation_overview}, developers initiate mitigation for 1089 of the 1359 vulnerabilities (80.1\%). For the remaining 270 vulnerabilities, we find no evidence of corrective work. Among these vulnerabilities, 149 receive no developer response, whereas 121 receive a response without observed mitigation. Of these 121 response-only reports, 43 are closed without verified mitigation.
These observations show that developer engagement and report closure alone do not establish effective remediation of the reported vulnerabilities.

To understand how these vulnerabilities reach closure without verified mitigation, we examine the associated developer discussions.
Our analysis reveals that some mitigation requirements depend on production configurations that cannot be verified from the repository.
In such cases, developers may acknowledge the required protections without confirming whether they are active in production.
For example, a vulnerability in \textsc{jevtrafficsim}\footnote{https://github.com/skcache/jevtrafficsim}’s policy API allows callers to bypass rate limits by forging the forwarding headers used to identify clients.
The developer documents code hardening and tests, but the required production firewall configuration remains unverified.
The issue is subsequently closed to consolidate mitigation work under a separate tracker, with the developer acknowledging the unresolved requirement:

\begin{table}[t]
\centering
\caption{Distribution of developer responses, mitigation strategies,
and mitigation effectiveness.}
\label{tab:mitigation_overview}
\small
\setlength{\tabcolsep}{3pt}
\renewcommand{\arraystretch}{1.05}
\begin{tabularx}{\columnwidth}{
    @{}>{\raggedright\arraybackslash}p{1.65cm}
    >{\raggedright\arraybackslash}X
    >{\raggedleft\arraybackslash}p{0.65cm}@{}
}
\toprule
\textbf{Category} & \textbf{Item} & \textbf{Count} \\
\midrule

\multirow[t]{3}{=}{\raggedright\textbf{Developer response}}
& No response & 149 \\
& Response only & 121 \\
& Mitigation initiated & 1089 \\

\midrule

\multirow[t]{6}{=}{\raggedright\textbf{Mitigation strategy}}
& Failure handling & 295 \\
& Decision constraints & 365 \\
& Data and credential hygiene & 414 \\
& Access control and resource limits & 414 \\
& Integration fixes and observability & 363 \\
& Security notices and feature disablement & 66 \\

\midrule

\multirow[t]{3}{=}{\raggedright\textbf{Mitigation effectiveness}}
& Exploitation blocked & 755 \\
& Residual exploitability & 198 \\
& Configuration-dependent exploitability & 136 \\

\bottomrule
\end{tabularx}
\end{table}

\begin{developerquote}{Developer Comment - \textsc{jevtrafficsim} Issue \#55}
\small
Important \#53 evidence: external probing of /.well-known/vercel/rate-limit-api/\textless{}ruleId\textgreater{} fell through to the app's own 404 page for both the real rule id and a bogus control id, so external probing cannot prove the platform rule is active. Dashboard or credentialed CLI verification remains
required.
\end{developerquote}

Closure can also follow a documentation update that leaves the affected code unchanged. For instance, an unrecognized judge mode in \textsc{need2know}\footnote{https://github.com/qfennessy/need2know} causes the system to replace the intended Jev access check with a permissive local check based on word overlap.
The corresponding pull request is merged without correcting the identified access control weakness, and the developer states:

\begin{developerquote}{Developer Comment - \textsc{need2know} Pull Request \#5}
\small
Known, documented but not fixed: N2K\_JUDGE\_MODE fails open: judge.py selects the live judge only on an exact jev, so any other
value silently selects the permissive local overlap judge. Now called out in the README; worth a separate code fix.
\end{developerquote}

\finding{8}{Developers initiate mitigation for 80.1\% (1089/1359) of the reported vulnerabilities. However, 44.8\% (121/270) of cases without observed corrective work received a developer response, and 35.5\% (43/121) of response-only reports are closed without verified mitigation. Developer responses and report closure should therefore not be treated as evidence of remediation.}

We next examine the security measures used to address the reported vulnerabilities. As shown in Table~\ref{tab:mitigation_overview}, developers apply six classes of mitigation strategies to the 1089 vulnerabilities for which mitigation is initiated.
These strategies are not mutually exclusive because more than one can be applied to the same vulnerability.
\textit{Data and credential hygiene} and \textit{access control and resource limits} are the two most frequently adopted strategies, each applied in 414 vulnerabilities (38.0\%), followed by \textit{decision constraints} with 365 vulnerabilities (33.5\%) and \textit{integration fixes and observability} with 363 vulnerabilities (33.3\%).
The prevalence of these two strategies aligns with the vulnerability landscape in RQ1, where V9 \textit{Unintended Sensitive Data Transmission} and V10 \textit{Private Application State Exposure} are the two most prevalent vulnerability types.
Measures under these two leading strategies include sensitive data filtering and redaction, limits on storage and propagation, validation of transmission destinations, enforcement of authentication and authorization requirements, and restrictions on service usage and resource consumption. These measures target sensitive information exposure, unauthorized service access, and uncontrolled resource usage in System One model applications.

\finding{9}{Developers generally employ six classes of mitigation strategies for the reported vulnerabilities in System One model applications. Among these, \textit{data and credential hygiene} and \textit{access control and resource limits} are the two most frequently adopted strategies, with measures targeting sensitive information exposure, unauthorized service access, and uncontrolled resource consumption.}

To assess mitigation effectiveness, we examine the extent to which the adopted strategies prevent exploitation of the reported vulnerabilities.
A mitigation is considered effective if it blocks the reported exploitation paths under the intended configuration.
As shown in Table~\ref{tab:mitigation_overview}, developers initiate mitigation for 1089 of the 1359 vulnerabilities. 
Of those receiving mitigation, 755 (69.3\%) have their reported exploitation paths blocked, and 136 (12.5\%) are exploitable only under improper configuration.
However, the remaining 198 vulnerabilities (18.2\%) retain exploitable paths after corrective efforts.
Combined with the 270 vulnerabilities without corrective action, 468 of the 1359 vulnerabilities (34.4\%) lack an effective resolution. 
Overall, mitigation efforts remain incomplete, with a substantial proportion of reported vulnerabilities lacking verified remediation and some remaining exploitable even after developer intervention.

\finding{10}{More than one third (34.4\%) of the collected vulnerabilities have not been effectively mitigated. Among the vulnerabilities for which developers initiated mitigation, 18.2\% remain exploitable despite corrective efforts. This suggests that the observed mitigation efforts remain insufficient to achieve comprehensive remediation of the reported vulnerabilities.}

We further examine the vulnerabilities that remain exploitable after mitigation.
Of the 334 vulnerabilities examined, 136 (40.7\%) are exploitable only in misconfigured deployments.
Among the adopted strategies, \textit{integration fixes and observability} is the most prevalent, applied to 73 vulnerabilities (53.7\%).
This strategy includes corrections to provider configurations, extensions of security checks to previously unchecked operations, and explicit error reporting when these checks cannot execute.
The effectiveness of these changes depends on proper deployment and enforcement of the required security controls, as misconfigurations may leave the reported exploitation paths open.
For the remaining 198 vulnerabilities (59.3\%), the adopted mitigations leave at least one reported exploitation path unblocked even under intended configurations.
These corrective efforts most commonly rely on \textit{data and credential hygiene}, accounting for 84 vulnerabilities (42.4\%).
Typical measures under this strategy include sensitive data filtering and redaction, restrictions on credential storage and propagation, and sanitization of logs and outputs.
Such protections cover only selected credential formats or data flows, leaving other disclosure paths unprotected.

\finding{11}{Current mitigation efforts often fall short of robust remediation. Some protections depend on correct configuration, while others leave exploitation paths reachable despite correct deployment. Effective fixes must therefore ensure reliable enforcement of security controls and address the underlying defects across all relevant execution paths.}

Our analysis further reveals that developers recognize the limitations of their mitigations for some vulnerabilities yet defer stricter enforcement to preserve application functionality.
This trade-off is particularly pronounced in applications that employ System One models as security controls, which account for 90 of the 198 vulnerabilities with residual exploitability (45.5\%).
In such applications, mitigation entails balancing security policy enforcement against the preservation of legitimate operations. 
Stricter measures may increase false rejections, while enforcing mandatory security checks may disrupt application workflows during model service failures.

An illustrative example is the mitigation of a vulnerability in the \textsc{super-jev}\footnote{https://github.com/Kevthetech143/super-jev} application, where the security gate misinterprets confidence in a \texttt{NOT\_SUPPORTED} verdict as evidence of support, resulting in a bypass of the intended claim validation mechanism.
The developer revises the handling of insufficient evidence but retains the vulnerable threshold logic due to concerns about introducing new false rejections:

\begin{developerquote}{Developer Comment - \textsc{super-jev} Pull Request \#18}
\small
I did not flip it in this PR. Making the gate stricter
without a fresh bench run would trade tonight's false
blocks for a new, unmeasured set of them, and this PR
is a precision fix. It is recorded as open decision 1
in docs/hooks.md, in a long comment on the thresholds,
and in NOT\_SUPPORTED\_DIRECTION\_NOTE, which
--explain prints on every run. Sir, this is the one
thing here that needs your call, and it should be
settled with a bench run, not an edit.
\end{developerquote}

Another example of deferred security enforcement involves withholding a stricter approval policy to preserve automated workflows.
In the \textsc{omp-classifier}\footnote{https://github.com/STRML/omp-classifier} application, the security gate initially evaluates script invocation commands without inspecting the scripts themselves.
The developer addresses this weakness by incorporating script contents into the security assessment. This mitigation verifies the script before approval but does not ensure its integrity at execution time, allowing a process running under the same user account to substitute unreviewed code after approval. 
The developer accepts this residual risk rather than enforce mandatory manual approval, which would prevent automated script execution:

\begin{developerquote}{Developer Comment - \textsc{omp-classifier} Issue \#128}
\small
Sam's worker fleets run scripts headlessly as
their normal mode of work, so that posture would block
the fleet rather than protect it. The three other P1s
from the review (per-verb option grammar, discarded
loader refusals, per-segment cd) are merged; this one
is not.
\end{developerquote}

\finding{12}{Residual exploitability can reflect deliberate acceptance of security risks rather than a lack of developer awareness. Developers may refrain from adopting stricter safeguards to limit false rejections or preserve automated execution. This suggests that the practical adoption of stronger mitigations in System One model applications depends not only on their security guarantees but also on their ability to preserve legitimate application functionality.}

\section{Discussion}

\subsection{Security Implications of Decision Delegation}
System One model applications introduce a programming paradigm in which developers delegate decisions that are difficult to express through conventional program logic to AI models. 
Such delegation, however, raises a fundamental security question concerning the extent to which model judgments should govern application behavior.
The structured answers and associated probabilities returned by System One models characterize the supplied context rather than establish that subsequent operations comply with application security policies. 
A model may correctly determine that an operation is benign without establishing that the requesting party is authorized to perform it. 
Similarly, high confidence in a judgment does not exempt an operation from mandatory security checks.
The security risks of decision delegation thus call for a distinction between decisions that benefit from probabilistic reasoning and security guarantees that must hold independently of model behavior.

One way to realize this distinction is to separate model judgments from the enforcement of mandatory security policies. Applications can use model judgments to guide decisions while independently validating the resulting operations before execution. This separation motivates several directions for security research, including static analysis to identify unintended dependencies on model judgments, dynamic testing via erroneous judgment injection, and formal verification of security properties that must hold regardless of model behavior. Such analyses could assess the security consequences of decision delegation independently of model reliability, enabling developers to determine which decisions can safely be delegated in the first place.

\subsection{Least Privilege for Judgment Context}

Decision delegation in System One model applications introduces a further security concern regarding the information disclosed to obtain model judgments.
An application may have legitimate access to sensitive information without authorization to disclose it to a model provider.
Context construction thus constitutes a security decision rather than a mere input preparation step, as it determines what application state is exposed to the model provider.
Without appropriate constraints on context selection, even requests to authorized providers may disclose sensitive information irrelevant to the delegated decision.
This risk motivates applying the principle of least privilege at the granularity of individual judgments rather than entire applications.

Enforcing least privilege for judgment context requires consideration of how context reduction affects judgment quality. A uniform redaction policy may remove evidence necessary for one judgment while retaining sensitive information irrelevant to another. 
This calls for techniques that reconcile judgment requirements with confidentiality constraints. 
To this end, information flow analysis could identify unauthorized disclosures, while context ablation could determine which data elements are necessary for reliable judgments.
Local abstraction could further assess whether reliable judgments require access to sensitive data in its original form.
Such techniques could make least privilege a testable property of individual judgment requests by assessing both the permissibility and necessity of information disclosure.

\subsection{The Cost of Secure Automation}

A key motivation for adopting System One models is to automate decisions that would otherwise require human intervention or complex program logic. Such decisions may govern operations with security consequences, such as information disclosure, resource access, and command execution. 
Preserving application security may therefore require additional validation, human review, or suspension of execution when valid judgments are unavailable.
Such safeguards can undermine the efficiency and autonomy that motivate decision delegation. 
As revealed in Finding 12, these operational costs may discourage developers from adopting stricter protections despite their awareness of residual security risks.
The cost of enforcement can thus constrain the security guarantees realized in practice, even when stronger safeguards are technically feasible.

Reducing these operational costs does not necessarily require relaxing security policies. An unresolved authorization decision for one operation should not impede other operations that can proceed securely within the same workflow. A promising research direction is to decouple the authorization of individual operations from the progress of the surrounding workflow. Analysis of operation dependencies and policy requirements could determine which operations can proceed while others await model judgments or human approval. This reframes secure automation as an optimization problem whose objective is to maximize autonomous execution under fixed security requirements.

\subsection{Systemic Risks of Shared Integrations}

The integration of System One models into applications often relies on reusable software artifacts, such as SDKs, integration libraries, and code examples. These artifacts not only facilitate model adoption but also provide common implementations for judgment interpretation, error handling, and decision enforcement.
For example, a shared SDK may return a low risk score when the model fails to provide a judgment. Applications that rely on this score for authorization may consequently permit operations without a valid security assessment. Defects in shared integrations can therefore compromise otherwise independent applications that reuse the affected code. 
The security significance of such defects depends on their local impact and the extent of code reuse across the ecosystem.

Shared integrations can create an asymmetry between the propagation of vulnerabilities and that of security fixes. Applications that import shared libraries may benefit from upstream fixes through dependency updates, whereas those that adopt code examples may retain vulnerable implementations even after the original defects are corrected. Moreover, identifying common defects through code similarity alone may be insufficient, as their security consequences depend on how individual applications use model judgments. 
Future research could derive reusable security specifications from reported vulnerabilities and their fixes to guide and validate remediation across related integrations. This could extend the reach of remediation beyond shared code to independently developed implementations that violate the same security properties.

\subsection{Toward System One Security Analysis}

Security analysis of System One model applications presents a distinct challenge because the semantics of application decisions are distributed across natural language judgment criteria and program logic. Typed model interfaces specify the structure of returned answers, but their meaning depends on the questions supplied to the model. A high probability can represent either a favorable or an unfavorable assessment of an operation, depending on the formulation of the judgment question. Conventional checks on response types and value ranges cannot detect incorrect interpretations of otherwise valid judgments. Similarly, evaluating the model in isolation may overlook discrepancies between the intended meaning of its judgments and their use in application logic. Detecting such vulnerabilities requires security analysis that bridges the semantic gap between judgment specifications and program behavior.

This semantic gap becomes particularly important as System One model applications evolve. A change to a judgment question may alter the security meaning of a model answer without changing its type or the application logic that consumes it. Such changes can invalidate existing security checks while preserving interface compatibility.
Addressing these semantic regressions opens three complementary research directions.
First, semantic versioning could identify revisions to judgment questions that require changes to security checks. Second, differential testing could detect unintended changes in application behavior across revisions of judgment specifications. Third, automated repair could reconcile revised judgment criteria with existing decision logic without weakening security policies.
These efforts could shift the focus of security regression analysis from individual code changes to the consistency between judgment specifications and their enforcement in application logic.

\section{Related Works}
\subsection{System One Model Security}
Since the release of Jev, studies have investigated the security of System One models.
Sun et al.~\cite{sun2026type} show that Jev's decisions can follow option names despite conflicting definitions.
Wu and Lim~\cite{wu2026decision} examine prompt injection against Jev and observe changes in action probabilities with limited success in inducing the attacker's intended choice.
Guo et al.~\cite{guo2026just} evaluate Jev as a detector of AI alignment failures, while Xu et al.~\cite{xu2609jevout} demonstrate decision manipulation through natural context.
Hu et al.~\cite{hu2026jevadvbench} introduce JevAdvBench and identify sensitivity to unverified opinions. 
Liu~\cite{liu2026evaluating} examines the reliability of System One models for automated security decisions.
Wang et al.~\cite{wang2026hidden} investigate security, privacy, and dual use through evaluations on Jev and NanoJev.
Azizi et al.~\cite{azizi2026one} expose guardrail bypasses through misleading option names. Li et al.~\cite{li2026same} compare Jev's adversarial robustness with other security judges, and Lan et al.~\cite{lan2026decisionrobust} introduce a robustness benchmark across textual and visual inputs. Yang et al.~\cite{yang2026tuple} investigate information leakage through typed outputs and probabilities. 
Kisielewicz et al.~\cite{kisielewicz2026escalate} examine the robustness of decision models in web attack triage and demonstrate that adaptive attacks can evade model judgments.
Lee et al.~\cite{lee2026jev} investigate probabilistic safety judges and identify limitations of confidence based escalation in detecting harmful outputs.
However, these studies primarily evaluate model behavior within predefined tasks and security assumptions. 
It remains unclear how vulnerabilities arise and are mitigated in real-world applications. Our work addresses this gap through a systematic measurement study of vulnerability characteristics, underlying causes, and mitigation practices in System One model applications.

\subsection{System One Model Application Security}
An emerging body of work examines the security of System One model applications.
Dong et al.~\cite{dong2026dice} demonstrate that manipulation of state extraction, question generation, or output interpretation can alter the final decision in a synthetic Jev pipeline despite authentic model responses.
Wu et al.~\cite{wu2026poisoned} trace context poisoning from retrieval to execution in two synthetic Jev workflows and examine the tradeoffs between attack prevention and legitimate automation.
Hu et al.~\cite{hu2026jev} integrate Jev-Guard into the Hermes agent framework to combine policy routing and safety classification with separate permission checks for agent actions.
Luo et al.~\cite{luo2026nudge} introduce JevProbe to evaluate decision hijacking and its potential security consequences in 228 open-source Jev applications, and also assess the effectiveness of existing defenses.
Shoham et al.~\cite{shoham2026think} introduce JevAct to verify tool actions in agent workflows through deterministic state checks and structured model judgments. 
Ghadamyari and Habibi~\cite{ghadamyari2026decision} evaluate decision models as retrieval gates against prompt injection and information pollution, and examine the effects of filter placement on downstream responses. 
Wang et al.~\cite{wang2026s1} introduce S1-Evo and demonstrate how adversarial manipulation of model judgments leads to security policy violations in agent workflows. 
While prior studies have examined specific security threats and defensive mechanisms, a critical gap remains in systematically understanding the broader vulnerability landscape and mitigation practices in System One model applications.
Therefore, we aim to inform secure application design and effective remediation through a systematic analysis of implementation weaknesses and practical constraints on mitigation.

\section{Conclusion}
In this paper, we present the first systematic measurement study of security vulnerabilities in System One model applications. By analyzing 1,359 vulnerabilities across 523 applications, we characterize the vulnerability landscape, identify underlying failure mechanisms throughout the application workflow, and assess the effectiveness of developer mitigations. Our findings demonstrate that reliable model judgments alone do not ensure secure application behavior and that mitigation effectiveness is constrained by incomplete security enforcement and the need to preserve application functionality. These observations underscore the need to treat security as an end-to-end property of model-mediated decision making rather than a property of the model alone. We hope our study provides an empirical foundation for effective security analysis and mitigation techniques and inspires further research toward securing the emerging ecosystem of System One model applications.

\bibliographystyle{ACM-Reference-Format}
\bibliography{sample-base}


\end{document}